\documentclass[12pt]{elsart}
\usepackage{natbib}
\usepackage{amsmath}
\usepackage{amsfonts}
\usepackage{amssymb}
\usepackage{epsfig}
\usepackage{epstopdf}
\usepackage{graphics}
\usepackage{epsf}
\newtheorem{theorem}{Theorem}

\newtheorem{definition}[theorem]{Definition}

\newtheorem{remark}[theorem]{Remark}

\begin{document}
\begin{frontmatter}
\title{A complete classification of epistatic two-locus models}
\journal{Genetics}
\date{}
\maketitle
\author{Ingileif B. Hallgr\'{\i}msd\'ottir\corauthref{cor}}
\address{Department of Statistics\\University of Oxford\\email: \tt{ingileif@stats.ox.ac.uk}}
\corauth[cor]{Corresponding author}
\author{Debbie S. Yuster}
\address{Department of Mathematics\\Columbia University\\email: \tt{debbie@math.columbia.edu}}

\begin{abstract}
  The study of epistasis is of great importance in statistical
  genetics in fields such as linkage and association analysis and QTL
  mapping.  In an effort to classify the types of epistasis in the
  case of two biallelic loci Li and Reich listed and described
  all models in the simplest case of 0/1 penetrance values.  However,
  they left open the problem of finding a classification of two-locus
  models with continuous penetrance values.  We provide a complete
  classification of biallelic two-locus models. In addition to solving
  the classification problem for dichotomous trait disease models, our
  results apply to any instance where real numbers are assigned to
  genotypes, and provide a complete framework for studying epistasis
  in QTL data.  Our approach is geometric and we show that there are
  387 distinct types of two-locus models, which can be reduced to 69
  when symmetry between loci and alleles is accounted for.  The model
  types are defined by 86 circuits, which are linear combinations of
  genotype values, each of which measures a fundamental unit of
  interaction.  The circuits provide information on epistasis beyond
  that contained in the $additive \times additive$, $additive \times
  dominance$, and $dominance \times dominance$ interaction terms.  We
  discuss the connection between our classification and standard
  epistatic models and demonstrate its utility by analyzing a
  previously published dataset.
\end{abstract}

\end{frontmatter}

\section{Introduction}
The genetic dissection of complex traits is at the center of current
research in human genetics.  Complex traits are caused by multiple
susceptibility genes and environmental factors, and mounting evidence
from both human genetics and model organisms suggests that
\emph{epistasis} (gene $\times$ gene interaction) plays an important
role~\citep{Clark1997,Storey2005}.
Although the need to consider epistasis when mapping complex trait
loci has been discussed by several authors
\citep{Carlborg2004b,Wade2001,Hoh2003,Moore2005}, most statistical methods used in gene
mapping, be it case-control association studies, quantitative trait
loci (QTL) mapping, or linkage analysis, are based only on measures of
marginal effects at individual loci and do not consider epistasis.
Due to recent advances in genotyping technology many large
case-control genome-wide association studies~\citep{Hirschhorn2002}
are underway, and there has been renewed interest in two-locus disease
models and two-locus tests for
association~\citep{Marchini2005,Evans2006,Zhao2006,Perez2006}.  The
application of two-locus models also arises in expression QTL mapping
where thousands of gene expression traits are mapped with linkage
analysis and it is imperative to study gene
interactions~\citep{Brem2005}.

Ideally, a test for epistasis between two loci $A$ and $B$ should test
for {\it biological interaction}, or non-independence between the
effects of locus $A$ and locus $B$. Loci $A$ and $B$ are considered
independent if the effect of the genotype at locus $A$ does not depend
on the genotype at locus $B$.
This biologically motivated concept has been formalized in a variety
of ways by different communities seeking simple, mathematically
convenient definitions.  In the statistical genetics literature the
term epistasis is typically taken to mean that the effects at loci $A$
and $B$ are not additive (the ``effect'' of a locus is defined in
terms of the statistical model used~\citep{Cordell2002}).  For a
further discussion on epistasis
see~\citet{Wade2002,Phillips1998,Routman1995}.  \citet{Fisher1918}
considered a linear model for the contribution of different loci to a
quantitative trait and used the term \emph{epistasy} to describe a
departure from additivity. In linkage analysis based on variance
component models, a model without epistasis is a model in which all
dominance variance components are zero.  In case-control association
studies of dichotomous traits it is common to use logistic regression,
and additivity is measured in the log-odds of disease for a
genotype~\citep{Gauderman2002}.  Recently~\citet{Zhao2006} suggested a
new test for epistasis which tests for departures from linkage
disequilibrium (LD) in the cases, which is equivalent to testing for
departure from additivity of the log penetrance values (i.e. departure
from a multiplicative model for the two-locus penetrances).  These
tests all test for departure from additivity on a particular scale,
but if an additive model is rejected they provide no information on
the type of interaction present. Furthermore, it is not clear what the
biological meaning of the interaction is.

With each of the nine two-locus genotypes we associate
a {\em genotype value}.  
In the case of a dichotomous phenotype the genotype value is the
penetrance associated with the genotype, and in the case of a
quantitative trait, the genotype value is the expected phenotype value
of individuals with that genotype (sometimes called {\em measured
  genotype}).  We will consider epistasis to be any deviation from
additivity of the genotype values.  This is consistent with the
definition of epistasis given in \citet{Cordell2002}, both for
quantitative and dichotomous traits.  In this paper we provide a
framework within which one can study and classify the types of
epistasis possible between two biallelic loci.  Our results are based
on recent work of \citet{Beerenwinkel2006} who provide a rigorous
geometric approach to epistasis in the haploid case. We extend their
results to the diploid case, and characterize all possible patterns of
physical interactions among the 9 possible genotypes in the two locus
case, showing that there are 387 classes of models that fall into 69
symmetry classes.  We discuss the meaning of the different types of
interaction and show how the interaction pattern can be effectively
measured and visualized.

In genetic analysis it is common to test not only for departure from
additivity, but also for whether the data fits a particular two-locus
model (e.g.\ recessive or dominant).  In Section~\ref{sec:models} we
discuss the models that are frequently used and show how they relate
to the classification given here.  In order to study a wider class of
two-locus models, ~\citet{LiReich2000} enumerated all two-locus,
two-allele, two-phenotype disease models with penetrance values 0 or 1
for the nine possible phenotypes.  There are 512 such models, which
can be reduced to a list of 50 models after allowing for symmetry
between alleles, loci and affection status. We classify models with
continuously varying penetrances, overcoming the difficulty they
highlight in their paper, and show that in fact their 50 models fall
into 29 of the 69 symmetry classes.

In Section~\ref{sec:shapes} we introduce the mathematical concepts
used to derive the 387 classes of two-locus models.  In Section
\ref{sec:classify} we demonstrate on a real dataset how the shapes can
be used to classify pairs of loci and identify pairs with similar
genetic effects.  In Section~\ref{sec:models} we consider the two-locus
models typically used in human genetics, the 50 models
from~\citet{LiReich2000}, and some models with epistasis. We show that
these models  only represent a small fraction of all possible
two-locus models.

\section{Shapes of two-locus models}
\label{sec:shapes}

A two-locus disease model on two biallelic loci is specified by the
genotype values of the 9 two-locus genotypes.  We consider two loci
with genotypes $aa, Aa,$ and $AA$, and $bb, Bb,$ and $BB$,
respectively, where $A$ and $B$ are the susceptibility alleles.  The
genotype values, $f_{ij}, i,j=0,1,2,$ are represented by a $3\times3$
table,

\begin{center}
\begin{tabular}{|l|ccc|}
\hline
     & $bb$ & $Bb$ & $BB$ \\
\hline
$aa$ & $f_{00}$ &  $f_{01}$ &  $f_{02}$ \\
$Aa$ & $f_{10}$ &  $f_{11}$ &  $f_{12}$ \\
$AA$ & $f_{20}$ &  $f_{21}$ &  $f_{22}$ \\
\hline
\end{tabular}
\end{center}

where $i$ and $j$ refer to the number of disease alleles at loci $A$
and $B$, respectively.
In the case of a dichotomous trait, $f_{ij}$ is a penetrance, or
probability that an individual with genotype $ij$ will get the
disease. For a quantitative trait, $f_{ij}$ is the expected phenotypic
value for an individual with genotype $ij$.

In an additive model, the genotype values can be written as a sum of
the effect at each locus, $f_{ij}=\alpha_i+\beta_j,$ where $\alpha_i$
is the effect associated with having $i$ disease alleles at the first
locus, and $\beta_j$ is the effect associated with having $j$ disease
alleles at the second locus.  An epistatic model is any non-additive
two-locus model.  To study epistasis we consider the interaction
space, which is the space of all two-locus models modulo the space
spanned by all additive two-locus models.  The interaction space is
spanned by a set of linear forms in the $\{f_{ij}\}$ called {\it
  circuits}.  There is a circuit for each set of 3 collinear points,
and for each set of four
points in the plane such that no three of the points are collinear. 
The coefficients in the linear form are such that the sum of the
points in the circuit, when scaled by these coefficients, is zero. For
example, the circuit arising from the points $f_{00}$, $f_{01}$,
$f_{20}$, and $f_{12}$ is
$$-3f_{00}+4f_{01}+f_{20}-2f_{12},$$ since
$$-3 \cdot (0,0)+4 \cdot (0,1)+(2,0)-2 \cdot (1,2)=0.$$
Every circuit with four points can be seen as a contrast between two
pairs of genotype values. For example, the above circuit is positive if
$4f_{01}+f_{20} \geq 3f_{00}+2f_{12}$ and negative otherwise.  There
are 86 circuits in total, each of which measures a specific deviation
from additivity.

To interpret the meaning of the circuits we first perform a change of
coordinates.  In quantitative trait genetics the phenotypic value is
often decomposed into additive ($f_a$ and $f_b$) and dominance
($\delta_a$ and $\delta_b$) main effects at loci $A$ and $B$
respectively, and four epistatic effects, $additive \times additive$
($I_{AA}$), $additive \times dominance$ ($I_{AD}$), $dominance \times
additive$ ($I_{DA}$), and $dominance \times dominance$ ($I_{DD}$)
effects.  We will use the same notation here to decompose the genotype
values into main and epistatic effects.  We write the two-locus model
as

\begin{center}
\begin{tabular}{|l|ccc|}
\hline
     & $aa$ & $Aa$ & $AA$ \\
\hline
$bb$ & $\tilde{f}-f_a -f_b +I_{AA}$  &  $\tilde{f}+\delta_a - f_b-I_{AD}$ &  $\tilde{f}+f_a - f_b - I_{AA}$ \\
$Bb$ & $\tilde{f}-f_a + \delta_b - I_{DA}$ &  $\tilde{f}+\delta_a + \delta_b + I_{DD}$ &  $\tilde{f}+f_a + \delta_b + I_{DA}$ \\
$BB$ & $\tilde{f}-f_a + f_b - I_{AA}$ &  $\tilde{f}+\delta_a + f_b + I_{AD}$ &  $\tilde{f}+f_a + f_b + I_{AA}$ \\
\hline
\end{tabular}
\end{center}

where

\vspace{-0.3cm}

\begin{eqnarray*}
4 \cdot \tilde{f} &=& \phantom{-} f_{00}+f_{02}+f_{20}+f_{22} \\
4 \cdot f_a &=& -f_{00} + f_{02} - f_{20} + f_{22} \\
4 \cdot f_b &=& -f_{00} - f_{02} + f_{20} + f_{22} \\
4 \cdot \delta_a &=& -f_{00} + 2f_{01} - f_{02} -f_{20} + 2f_{21} - f_{22} \\
4 \cdot \delta_b &=& -f_{00} + 2f_{10} - f_{20} -f_{02} + 2f_{12} - f_{22} \\
4 \cdot I_{AA} &=& \phantom{-} f_{00} - f_{02} - f_{20} + f_{22} \\
4 \cdot I_{AD} &=& \phantom{-} f_{00} -2f_{01} +f_{02} -f_{20} + 2f_{21} - f_{22} \\
4 \cdot I_{DA} &=& \phantom{-} f_{00} -2f_{10} +f_{20} -f_{02} + 2f_{12} - f_{22} \\
4 \cdot I_{DD} &=& \phantom{-} f_{00} -2f_{01} + f_{02} -2f_{10} + 4f_{11} -2f_{12} +f_{20}-2f_{21}+f_{22}
\end{eqnarray*}

Note that with this choice the additive effect is scaled so that the
contribution is $-f_a$, $0$, and $f_a$ for genotypes $aa$, $Aa$, and
$AA$ respectively, and similarly for the second locus.  This is a
simple linear transformation of the genotype values which can be used
both for penetrances and phenotypic means. The space of all two-locus
models has dimension 9 and the interaction space has dimension 6.  A
natural choice of a basis for the interaction space is given by the
\emph{interaction coordinates} $(\delta_a, \delta_b, I_{AA}, I_{AD},
I_{DA}, I_{DD})$ where $\delta_a$ and $\delta_b$ measure within-locus
interaction and $I_{AA}, I_{AD}, I_{DA},$ and $I_{DD}$ measure
between-loci interaction.  A full list of the 86 circuits in the new
coordinates is given in Appendix~\ref{app:circ}.  
Note that the
circuits are a function of the interaction coordinates only.  The
circuits can now be interpreted, e.g.\ circuit
$c_{30}=2\delta_a-2\delta_b$ measures the difference between the
dominance effects,
circuit $c_1=-2\delta_a + 2I_{AD}$ measures the difference between the
dominance effect at the first locus and the $additive \times
dominance$ interaction, etc.  Although the circuits are fully
specified by the six interaction coordinates, they do contain
important information on the type of interaction present, and fully
describe all types of interaction.  The sign of a circuit specifies
whether the type of epistasis measured by the circuit is positive or
negative, and its magnitude measures the degree of interaction.  To
fully describe the pattern of interaction in a model, we can consider
the sign pattern of all 86 circuits.  The sign pattern gives
information beyond just considering the values of the six interaction
coordinates.

To classify all two-locus models according to the type of interaction
present, we consider the \emph{triangulation} induced by the
penetrances.  The mathematical definition of a triangulation is given
in Appendix~\ref{app:tri} 
but an informal description is provided
here.  We represent the 9 genotypes by 9 points in the plane on a $3
\times 3$ grid and the genotypic values by heights above these points.
If the values come from an additive model it is possible to fit a
plane through the height points.  For any non-additive model we
consider the surface given by the upper faces of the convex hull of
the heights.  Intuitively this is the surface formed if we were to
drape a piece of stiff cloth on top of the heights and consider its
shape.  Any departure from additivity in the model becomes apparent in
this surface.  The triangulation, or \emph{shape}, of a model is
obtained by projecting these upper faces (the ``creases'' in the
surface) onto the $xy$-plane.

A visual representation of a two-locus model is given in
Figure~\ref{fig:4panel}.  The data comes from an example that will
be discussed further in Section~\ref{sec:classify}.  The genotype
values, relative to the value of $aa/BB$, are listed in Panel (a).
Panel (b) shows the classical visualization of this table, where each
line corresponds to one row in the table.  In Panel (c) there is a
bar-chart of the data, and the corresponding shape is in Panel (d).
There is clearly epistatic interaction in the model in
Figure~\ref{fig:4panel}, as the genotypes $aa/bb$, $aa/Bb$, $Aa/bb$,
and $AA/BB$ have much higher means than the remaining 5 genotypes.
The shape shows the four planes of the upper convex hull of the
heights.  It includes a plane through the genotypes $Aa/bb$, $aa/Bb$,
and $AA/BB$, which is given by the middle triangle in the picture, and
three planes corresponding to the outer three triangles.

Although the classical visualization in Panel (b) contains complete
information on the relative genotype values, it is hard to grasp what
types of interactions occur just by glancing at the figure.  The
bar-chart is a very good visual representation of the 9 values,
however, any comparison between two different datasets based on
bar-charts would be not only tedious, but hard to define.  Some
information is lost by considering only the shape of the model, but
since it summarizes the epistasis that is present, the shape enables
us to easily compare and classify different models.

\begin{figure}
\begin{center}
\includegraphics[scale=0.55]{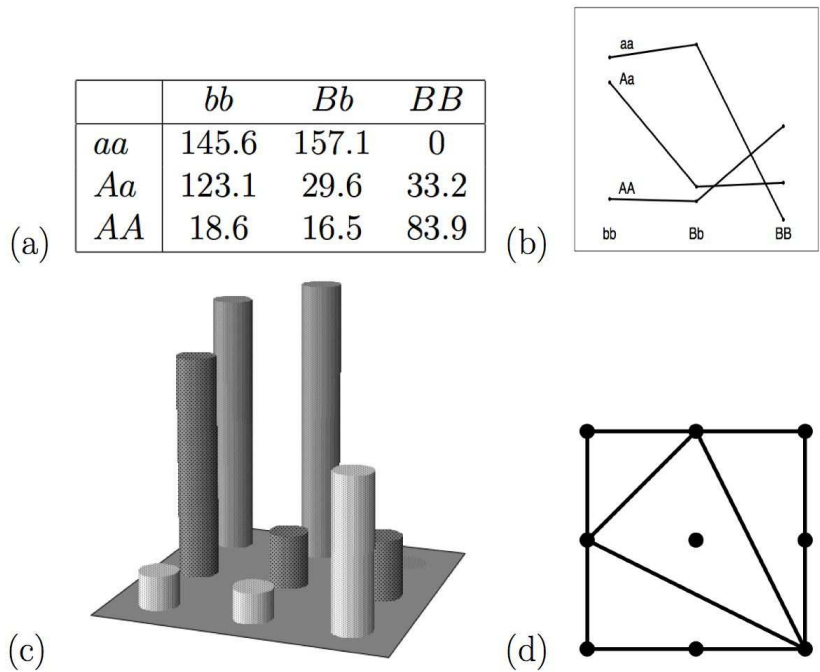}
\end{center}
\caption{Example of epistasis in QTL data.  The data is on chicken
  growth~\citep{Carlborg2004}.  (a) The phenotypic means of the
  two-locus genotypes, (b) a wiggle plot of the data, where each line
  corresponds to a row in the table, (c) bar plot of the data, (d) the
  two-locus shape.}
\end{figure}

\begin{figure}
\begin{center}
\includegraphics[scale=0.65]{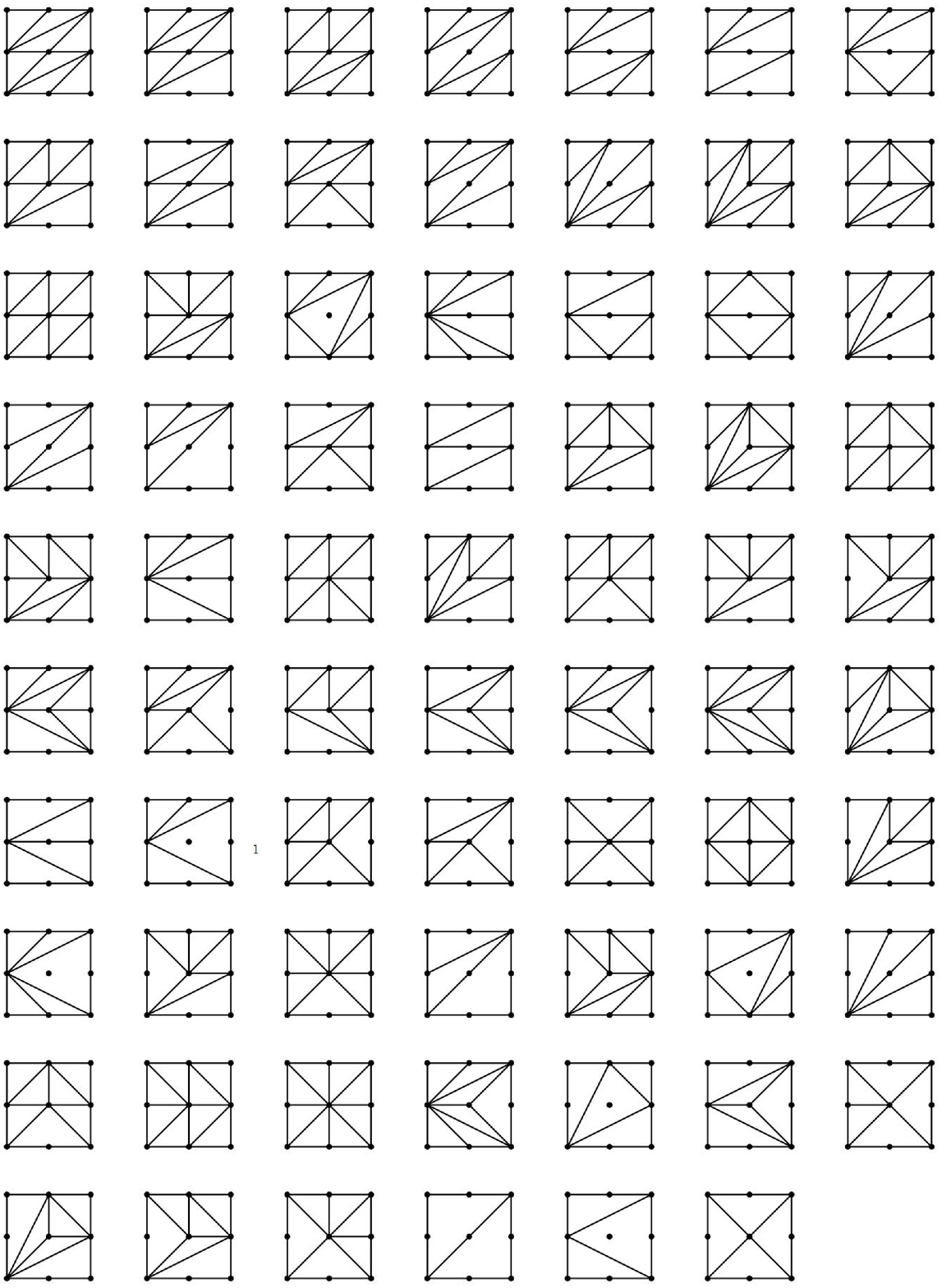}
\caption{The 69 symmetry classes of the shapes of two-locus models.}
\label{fig:69triangs}
\end{center}
\end{figure}

We used \begin{tt}TOPCOM\end{tt}~\citep{Rambau2002} to compute all
possible triangulations, or shapes, and found that there are 387,
however, many are equivalent when we account for symmetry. By symmetry
we mean i) the interchange of locus 1 and locus 2, or ii) the
interchange of two alleles at one or both loci.  These same symmetry
conditions were used in~\citet{LiReich2000}. After accounting for
symmetry, there are 69 shapes (see Figure~\ref{fig:69triangs}).  We
classify all two-locus models according to which of the 387 (or 69)
triangulations they belong to.

A sign pattern for the circuits specifies a model shape, but the
converse is not true. Thus considering the shape of a model, rather
than the sign pattern of the 86 circuits, gives a coarser model
classification, but it provides a very useful description of the type
of epistasis in the model.  A shape contains information about the
signs of some of the 86 circuits.  Every group of points in a circuit
can be triangulated in exactly two ways~\citep{GKZ} corresponding to
the type of epistasis.  If a model shape has a line connecting the
points $(i_1, j_1)$ and $(i_2, j_2)$ then for some circuit, $c=(a_1
f_{i_1j_1} + a_2 f_{i_2j_2}) - (b_1 f_{k_1l_1} + b_2 f_{k_2l_2})$, the
pair $f_{i_1j_1}$ and $f_{i_2j_2}$ are the ``winners'', i.e. $a_1
f_{i_1j_1} + a_2 f_{i_2j_2} \geq b_1 f_{k_1l_1} + b_2 f_{k_2l_2}$.
Similarly, if there is no line connecting the points $(i_1,j_1)$ and
$(i_2,j_2)$, and it is not possible to add one without crossing an
existing line segment, then there is some circuit such that
$f_{i_1j_1}$ and $f_{i_2j_2}$ are the ``losers''.  For example, in
Figure~\ref{fig:4panel}, there is a line between $(1,0)$ and $(0,1)$
and $f_{01}+f_{10} \geq f_{00} + f_{11}$, and also $2f_{01}+2f_{10}
\geq 3f_{00} + f_{22}$.  Note that the model shape gives information
about the types of interaction present in the model, but does not
reveal the magnitude of the interaction (for that we need the actual
value of the circuits).  For generic models we always get a
triangulation of the $3 \times 3$ grid, but for some models the
resulting shape is a subdivision, where not all cells in the shape are
3-sided (this happens e.g.\ when many of the genotype values are
identical).  These coarse subdivisions are not counted in our 387
models, however each coarse subdivision is refined by two or more of
our models. Looking at a specific triangulation or subdivision tells
us which way some (but not all) of the circuits are triangulated, thus
giving information about gene interaction for that particular
model. This is more informative than simply knowing whether the
interaction coordinates $I_{AA}$, $I_{AD}$, $I_{DA}$, and $I_{DD}$ are
zero or not.

Two shapes are adjacent if one can move from one to the other by
changing the sign of one of the 86 circuits.  Out of 387 shapes, 350
are adjacent to 6 other shapes, 16 are adjacent to 7 other shapes, and 21 are
adjacent to 8 other shapes.  We define the distance between two shapes
as the minimum number of circuit changes that are necessary to get
from one to the other.  In the set of 387 shapes the maximum distance
between two shapes is 9, and around 70\% of all pairs of shapes are
distance 4 to 6 apart.  Two-locus models which fall into adjacent
model shapes share many of the same two-locus interactions, and in
general the shorter the distance between two shapes, the more similar
the genetic effects.

Each shape divides the $3 \times 3$ grid into 2 to 8 triangles (the
numbers in each category are 2, 11, 38, 68, 96, 108 and 64 out of 387).
Each shape corresponds to a subspace of the 9-dimensional hypercube
and the volume of this subspace measures how much of the parameter
space the shape inhabits.  Shapes which divide the $3 \times 3$ grid
into fewer triangles occupy a larger part of this space.  We obtained
an estimate of this by generating 1,000,000 random vectors of length 9
and calculating the shape that each of them falls into.  The fraction
of observations that fell into shapes which divide the grid into 2, 3,
4, triangles, etc., was 6.4\%, 17.2\%, 28.3\%, 24.9\%, 15.0\%, 6.1\%
and 2.0\%, very different from the fraction of shapes in each category,
which is 0.5\%, 2.8\%, 9.8\%, 17.6\%, 24.9\%, 29.9\% and 16.5\%.
Two-locus models where one, or a few, genotype values are larger than
the remaining values induce shapes which contain fewer
triangles.  However, if the genotype values show only slight
deviations from falling on a plane (i.e.\ $\delta_a, \delta_b, I_{AA},
I_{AD}, I_{DA},$ and $I_{DD}$ are small), the surface is not dominated
by a few genotypes and resulting shape will be more subdivided.

In Section~\ref{sec:classify} we will further discuss how the shapes
can be used to characterize the type of interaction in a dataset.

\section{Two-locus models}
\label{sec:models}

In this section we study a number of model classes that are often used
in genetic analysis, and the shapes that they induce.  We show that
each of the model classes restricts the analysis to a small subset of
all possible two-locus models. Furthermore, because these models are
very specific, they limit the types of interaction that can be modeled
and only represent a small fraction of the 69 shapes.

A two-locus penetrance model can be defined by specifying single locus
penetrance factors, $(\alpha_0, \alpha_1, \alpha_2)$ and $(\beta_0,
\beta_1, \beta_2)$, and combining them in one of three ways,
\begin{center}
\begin{tabular}{rcl}
{\it multiplicative} &:& $f_{ij} = \alpha_i  \cdot  \beta_j$,  \\[-2mm]
{\it additive}       &:& $f_{ij} = \min(\alpha_i + \beta_j, 1)$,\\[-2mm]
{\it heterogeneous}  &:& $f_{ij} = \alpha_i + \beta_j - \alpha_i  \cdot  \beta_j$.
\end{tabular}
\end{center}
The penetrance factors are typically chosen from a recessive $(0, 0,
\alpha)$, dominant $(0, \alpha, \alpha)$ or additive $(0, \alpha/2,
\alpha)$ model.  For an additive two-locus model with additive
penetrance factors, the interaction coordinates $\delta_a, \delta_b,
I_{AA}, I_{AD}, I_{DA}, I_{DD}$ are all zero and the circuits all
vanish.  For all additive two-locus models $I_{AA}\!=\!I_{AD}\!=\!I_{DA}\!=
\!I_{DD}\!=0$, but $\delta_a$ and $\delta_b$ depend on the penetrance
factors.  The heterogeneous model is often viewed as an approximation
to the additive model because if the same penetrance factors are used,
the models give very similar penetrances.  However, in terms of the
type of interaction that can be modeled, the multiplicative and
heterogeneous models are very similar.  In Table~\ref{tab:intval}, we
list the values of the interaction coordinates for some common
multiplicative models.  If we consider the corresponding heterogeneous
models, the single locus dominance terms, $\delta_a$ and $\delta_b$,
have the same value, as listed in the table, and the interaction terms,
$I_{AA}$, $I_{AD}$, $I_{DA}$, and $I_{DD}$, all have the same absolute
value but opposite sign.  The shapes induced by these models are shown
in Figure~\ref{fig:twoloc_shapes}.  Note that 6 of the 8 shapes are
not generic models (they are subdivisions rather than triangulations
of the $3 \times 3$ grid).
 
\begin{center}
\begin{table}[ht]
\begin{center}
\begin{tabular}{|l|cccccc|}
\hline
One-loc  & $\delta_a$ & $\delta_b$ & $I_{AA}$ & $I_{AD}$ & $I_{DA}$  & $I_{DD}$\\
\hline
{\em rec-rec} & $-\eta_1/4$ & $-\eta_2/4$ & $\gamma$ & $-\gamma$ &  $-\gamma$ &  $\gamma$ \\
{\em rec-add} & 0 & $-\eta_2$ &  $\gamma$ & 0 & 0 &  -$\gamma$ \\
{\em rec-dom} & $\phantom{-} \eta_1/4$ & $\phantom{-} \eta_2/4$ & $\gamma$ & $\gamma$ &  -$\gamma$ &  -$\gamma$ \\
{\em dom-dom} & $\phantom{-} \eta_1/4$ & $\phantom{-} \eta_2/4$ & $\gamma$ & $\gamma$ &  $\gamma$ &  $\gamma$ \\
{\em dom-add} &  0 & $\phantom{-} \eta_2$ & $\gamma$ & $\gamma$ & 0 & 0 \\
{\em add-add} & 0 & 0 & $\gamma$ & 0 & 0 & 0 \\
\hline
\end{tabular}
\caption{The table lists the values of the interaction coordinates for
  multiplicative two-locus models.  The parameters are
  $\gamma = (\alpha_2-\alpha_0)(\beta_2-\beta_0), \eta_1 =
  (\alpha_0+\alpha_2)(\beta_2-\beta_0),$ and $\eta_2 =
  (\alpha_2-\alpha_0)(\beta_0+\beta_2)$.}
  \label{tab:intval}
\end{center}
\end{table}
\end{center}

\begin{figure}[h]
\label{fig:twoloc_shapes.eps}
\begin{center}
\includegraphics[scale=0.55]{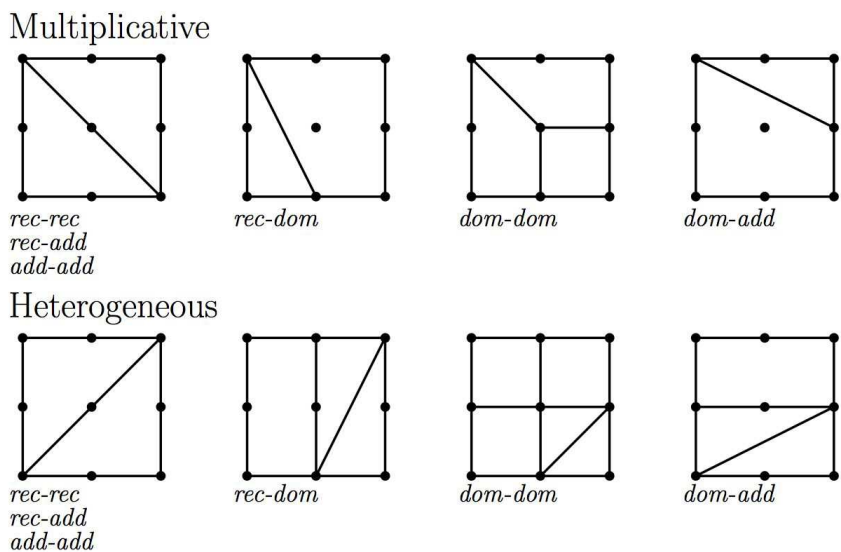}
\caption{The model shapes for multiplicative and heterogeneous two-locus models.}
\end{center}
\end{figure}

\begin{figure}[h]
\begin{center}
\includegraphics[scale=0.65]{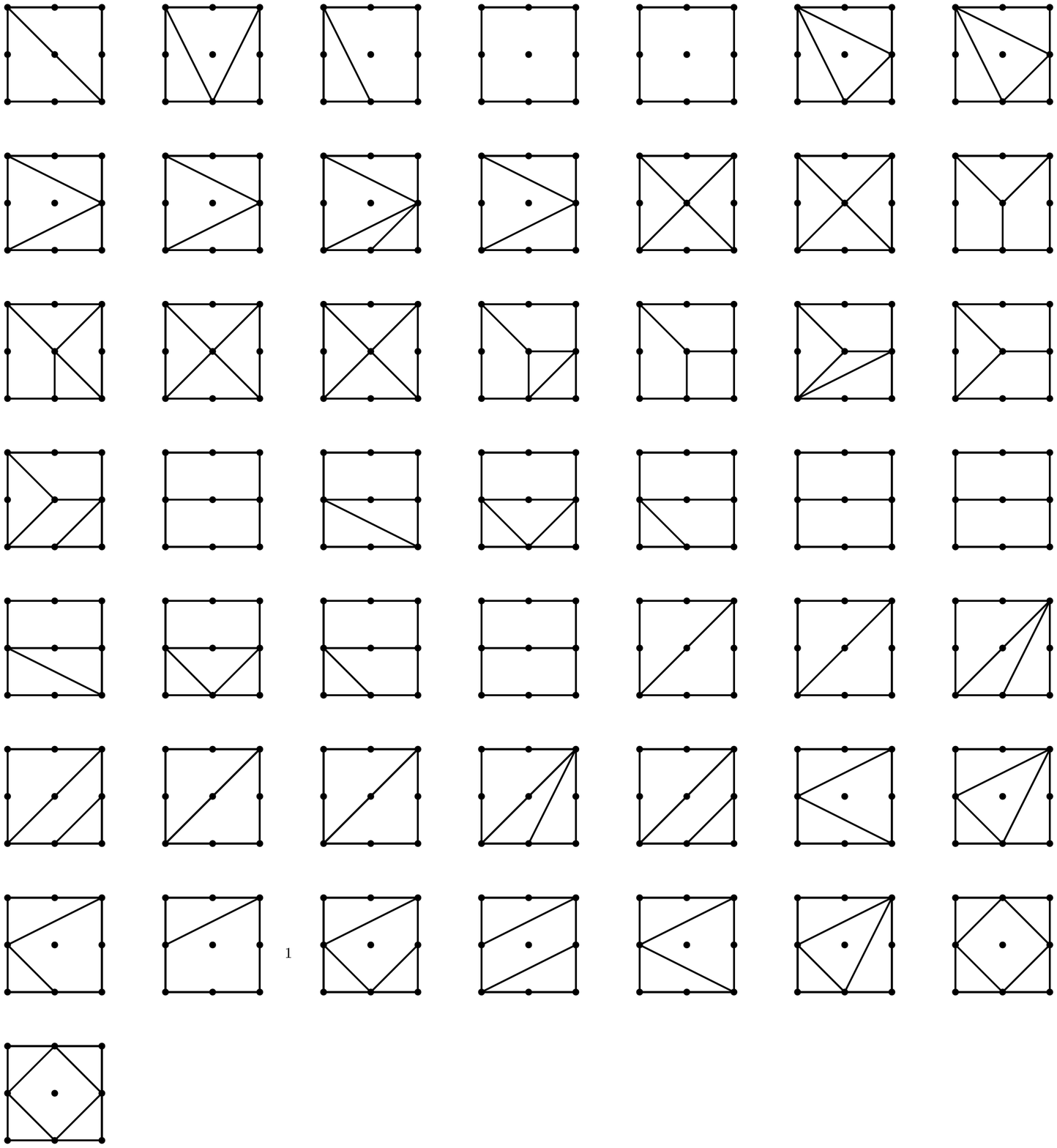}
\caption{The subdivisions for the 50 Li and Reich 0/1 penetrance models.}
\label{fig:50subdiv}
\end{center}
\end{figure}

~\citet{LiReich2000} described a classification of all two-locus
disease models with 0/1 penetrance values.  Although this
classification is useful to generate data under various scenarios and
to study general properties of two-locus models, it cannot be used to
classify observed data. This class of models is much larger than the
class of disease models discussed above, yet they only cover a small
part of all two-locus models.  The 50 models represent only 29 unique
subdivisions, and only 10 out of those 29 are among the 69 model
shapes, see Figure~\ref{fig:50subdiv}.

\begin{figure}[ht]
\begin{center}
\includegraphics[scale=0.55]{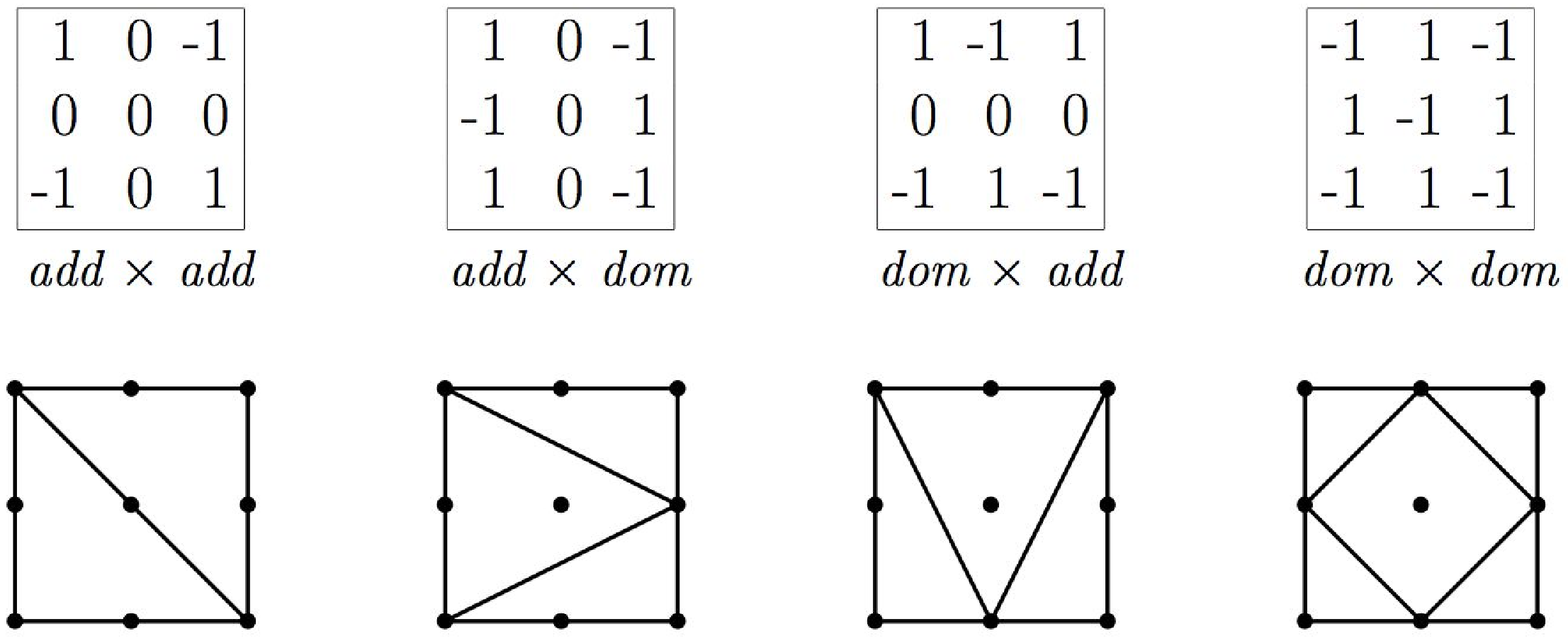}
\caption{The tables list the genotype values associated with four epistatic models and below each table is the shape induced by a model with purely $add \times add$, $add \times dom$, $dom \times add$ or $dom \times dom$ interaction.}
\label{fig:epishapes}
\end{center}
\end{figure}

In population genetics and in the study of quantitative traits,
two-locus models are classified according to the type of epistatic
effects.  Four commonly studied patterns of epistasis are {\em
  additive $\times$ additive}, {\em additive $\times$ dominance}, {\em
  dominance $\times$ additive} and {\em dominance $\times$ dominance}.
In an {\em additive $\times$ additive} model two double homozygotes,
$aa/bb$ and $AA/BB$, have higher phenotypic mean (or fitness) than
expected, but the other two, $aa/BB$ and $AA/bb$, have lower phenotypic mean than expected.
A numeric representation of the four types is given in
Figure~\ref{fig:epishapes} and the corresponding shapes are also shown.
If these epistatic effects are added to a fully additive two-locus
model, the resulting shape will be the one shown in
Figure~\ref{fig:epishapes}.  However, the epistasis observed in real data is seldom purely of one type. Furthermore, although e.g.\  {\em
  dominance $\times$ dominance} epistasis is present in the data, the
resulting shape can be different.  If the dominance terms, $\delta_a$
and $\delta_b$, are non-zero, the resulting shape will be the {\em
  dominance $\times$ dominance} shape, with the possible addition of
one or both of the horizontal and vertical lines through the middle of
the shape (depending on the magnitude of the dominance terms).  A model
with both {\em additive $\times$ dominance} and {\em dominance
  $\times$ additive} interaction can fall into one of three shapes.
If either the {\em additive $\times$ dominance} or the {\em dominance
  $\times$ additive} interaction is much stronger than the other, the
corresponding shape will dominate.  If the magnitude of both types of
interaction is similar, the resulting shape will be a shape shown in
Figure~\ref{fig:4panel} or any rotation thereof.  Thus from the shape we can often
infer what type of interaction is the strongest in the
data.

\section{Classification of epistatic effects}
\label{sec:classify}

Model organisms such as yeast, mouse or chicken are frequently used in
genetic analysis, and several recent studies have shown that epistatic
effects contribute greatly to observed genetic variability. 
When pairs of interacting loci have been found, using either QTL
mapping, linkage analysis, or association analysis, it is of interest
to describe the epistasis in the data. If many pairs of
interacting loci have been found, it is of interest to identify pairs
with similar genetic effects.  This classification can be based on
finding, for each pair, the model which best fits the data, out of the
models discussed in Section~\ref{sec:models}.  However, many datasets
do not fall into any one of these classes (e.g.\ more then one type of
epistasis can be present in the data).  Another option is to base the
classification on visual inspection, but that can be inaccurate and very time consuming, especially since in most applications the two alleles at a locus
are interchangeable, so one would have to consider many rotations of
the $3 \times 3$ data matrix.

We propose classifying observations according to the shape that they
induce, and measuring the similarity of the genetic effects observed
in two different datasets by the minimum distance between their induced shapes
(as defined in Section~\ref{sec:shapes}).  This
allows us to quickly and automatically identify observations with
similar genetic effects.  Beyond classification, our framework allows
us to use the values of the circuits to identify which interactions
are present in the data, and to test for specific interactions.  Tests
for specific interactions will give important information about the
true model shape, since there is always measurement error in
observations.  A more robust classification can be obtained by testing
which circuits are non-zero and considering the shape induced after
circuits which are not significant have been set to zero.

In a study of growth traits in chickens, \citet{Carlborg2004} measured
various growth and body weight variables on 546 chickens from an $F_2$
cross between two lines, a commercial broiler sire line and a White
Leghorn line.  The alleles at each locus are labeled with $B$ and
$L$, according to which line they came from.  \citet{Carlborg2004} used
a method for simultaneous mapping of interacting QTLs to do a
genome-wide analysis of five growth traits and identified 21 QTL pairs
with a significant genetic effect.  Some of the 21 QTL pairs were
associated with more than one growth trait, resulting in 30
combinations of traits and QTL pairs.  For each trait and QTL pair the
phenotypic means of each of the nine two-locus genotypes were
estimated using linear regression (see Table 2 in~\citet{Carlborg2004}).  They noted that the standard
models for epistasis do not adequately describe the types of
interaction present in their data, and classified the QTL pairs into
groups with similar genetic effect by visual inspection. They
identified 4 general classes of models in this dataset, and classified
16 out of the 21 QTL pairs into one of these classes (when a QTL pair
was associated with more than one trait the observations from both
traits were considered to be in the same class).  The classes are H)
some of the homozygote/heterozygote combinations are lower than
expected, B) the phenotype value associated with the genotype BB/BB is
lower than expected, A) the data fits an additive model, by visual
inspection, L) there is a set of genotypes with a high value, a set
with a low value associated with it, and the value associated with the
genotype $LL/LL$ is between the two, and U) the 5 QTL
pairs which did not fit into any of the four classes were left
unclassified.

\begin{figure}[ht]
\begin{center}
\hspace{-2cm} \includegraphics[scale=0.80]{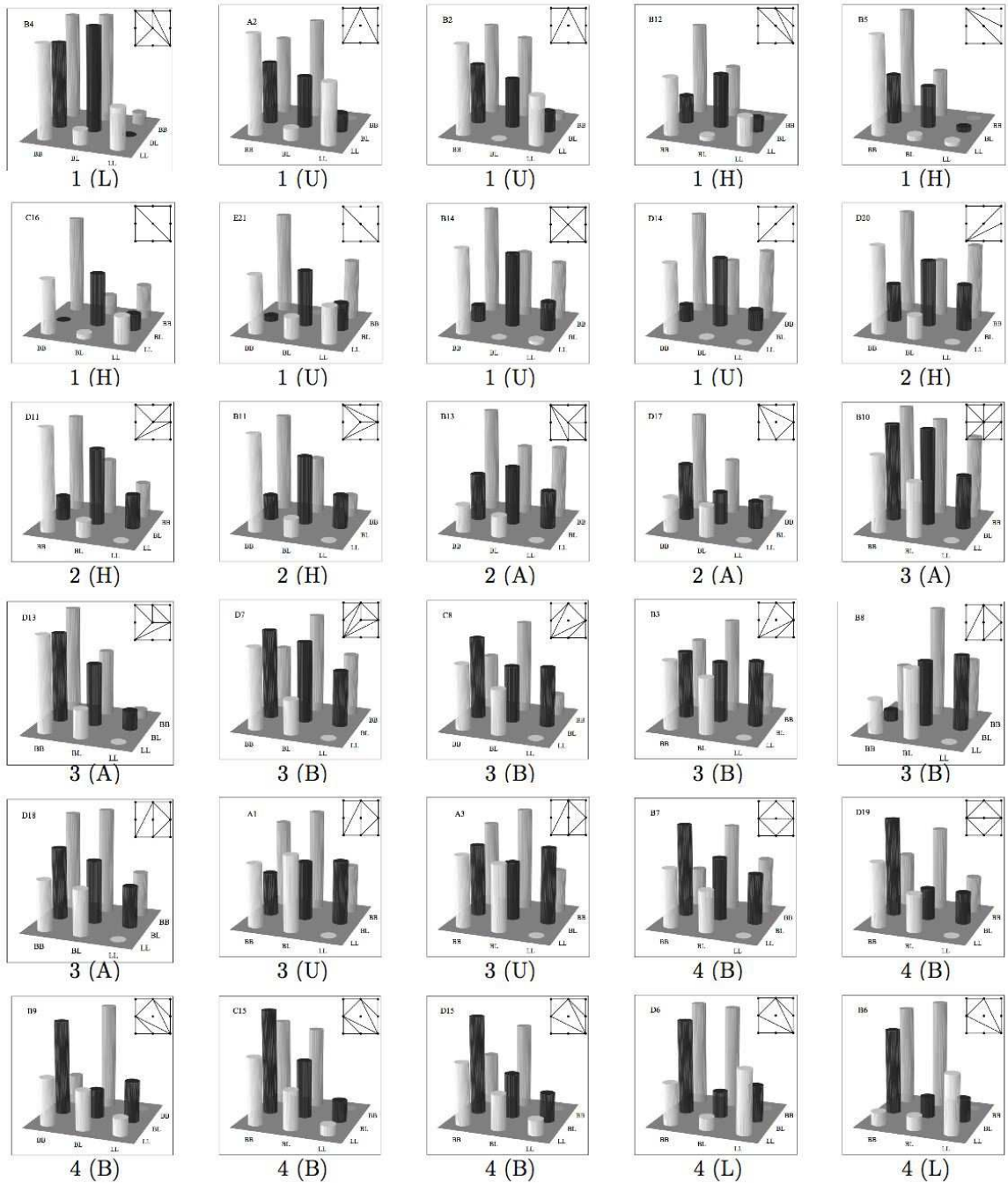}
\caption{A visual representation of the 30 trait/QTL pairs.  The
  phenotype average for each genotype is given by the heights of the
  bars, the corresponding shape is also given, and the trait (A-E) and
  QTL pair (1-21) listed.  Under each panel we list the cluster it
  falls into (1-4) and the group given by~\citet{Carlborg2004} (A, B,
  H, L, U).}
\label{fig:carlborg_data}
\end{center}
\end{figure}

We computed the shapes of the 30 observations and found that 23 of the
387 shapes occurred, or 16 out of 69 up to symmetry.  The data are
shown in Figure~\ref{fig:carlborg_data}.  For each observation we show
a bar-chart of the phenotype means and the corresponding shape.  The
point in the upper left corner of the shape corresponds to the
genotype $BB/BB$, and the point in the lower right corner corresponds
to $LL/LL$.  Although in most applications one would consider the two
alleles at a locus to be interchangeable we do not here, since they
come from different chicken lines.  To group together observations
with similar genetic effects we clustered the shapes based on the
pair-wise distances between them, using complete linkage hierarchical
clustering.  There are four main clusters in the resulting dendrogram
(not shown). Under each panel in Figure~\ref{fig:carlborg_data} we
list which cluster it falls into, and in parentheses we list which
group it belongs to according to \citet{Carlborg2004}.  The
observations are ordered based on the hierarchical clustering with
observations in the same cluster listed together and observations
within each cluster listed according to the distance between them, as
far as possible.  For four observations we switched the order of the
first and second locus, compared to the order in~\citet{Carlborg2004},
in order to minimize the distance to the closest observation.  Within
a cluster, the distance between the shapes in side-by-side panels is
typically one but occasionally two.  Many of the observed shapes are
adjacent to more than one other shape, so two shapes that are not
adjacent in Figure~\ref{fig:carlborg_data} may still be close.  

Consider the last row in Figure~\ref{fig:carlborg_data}.  In all five
panels the values of the genotypes $BB/BL$, $BL/BB$ and $LL/LL$
dominate the shape, resulting in a central triangular plane.  The
value at $BB/BB$ varies considerably but does not affect the shape.
The shape that each of the observations fall into is, however,
affected by the values of $BL/LL$ and $LL/BL$.  When they are 
relatively high an additional partition is added in the shape.
Recall from the previous section that this shape is observed when
there is both {\em additive $\times$ dominance} and {\em dominance
  $\times$ additive} interaction in the data.  The shapes in the
second-to-last row indicate strong {\em dominance $\times$ dominance}
interaction (compare to the shape given in Section~\ref{sec:models}).
In the last two observations in the row, {\em dom $\times$ dom} is the
strongest interaction, whereas the first three also show strong {\em
  add $\times$ dom} interaction.

The visual classification corresponds very well to the classification
based on shapes.  All observations labeled H fall into clusters 1 and
2 (which are close to each other in the dendrogram) and all
observations labeled B fall into clusters 3 and 4.  The observations
in group A (additive model) fall into two different clusters.  An
additive model has no shape (one can fit a plane through the points)
but due to measurement error in real data this will not be the case.
Note that 3 of the 5 observations in group A induce shapes which are
very subdivided, as can be expected when there are no genotypes with
very high values which dominate the shape.  The observations in group
U, which were previously unclassified, have now been grouped with the
observations they are closest to.  Two QTL pairs (4 and 6) were
grouped together in category L.  The two observations on QTL pair 6
are in cluster 4 and the observation on QTL pair 4 in cluster 1.

In case-control association studies the penetrances of the genotypes
are not observed, only the counts of cases and controls that have each
genotype.  To study the shape of a two-locus disease model we can fit
a full two-locus model using logistic regression (see
Section~\ref{sec:test}) and get the fitted log odds-ratio for each
genotype, which can then be used to obtain an estimate of the
penetrances.

\section{The power to detect epistasis depends on the model shape}
\label{sec:test}
The first step in a two-locus analysis is typically to identify
pairs of loci with statistically significant interaction, and then to
study the interaction.  The power to detect interacting loci depends
not only on the true model shape, but also on the minor allele
frequency at the two loci, the sample size in the study, the number of
genotyped markers, and the prevalence of disease/phenotypic mean in
the population.  Disregarding those other factors we ask how, or
whether, the power to detect interaction depends on the true model
shape.  To fully answer that question it is necessary to perform a
thorough simulation study which is outside the scope of this paper,
but we have performed a preliminary analysis with the goal of comparing
the relative power to detect interaction under different model shapes.
We considered three different situations: QTL mapping, association
analysis using logistic regression, and association analysis using an
LD based measure for interaction.

In two-locus QTL mapping, the phenotype is typically modeled as a
function of the genotype using a linear model. 
If $y$ is the phenotype, the model is
 \begin{eqnarray*}
 y &=&  \tilde{f} + f_a \cdot x_{A} + f_b \cdot x_{B} + \delta_a \cdot x_{Aa} + \delta_b \cdot x_{Bb} \\
   & & + I_{AA} \cdot x_{A}x_{B} + I_{AD} \cdot x_{Aa}x_{B}  + I_{DA} \cdot x_{A}x_{Bb} + I_{DD} \cdot x_{Aa}x_{BB} + \epsilon,
 \end{eqnarray*}
 where the coefficients of the model are the coordinates $\tilde{f}$,
 $f_a$, $f_b$, $\delta_a$, $\delta_b$, $I_{AA}$, $I_{AD}$, $I_{DA}$
 and $I_{DD}$, and $\epsilon$ is Gaussian. The $x_*$ are dummy
 variables; $x_{A}$ takes the values -1, 0, and 1 for individuals with
 genotypes $aa$, $Aa$, and $AA$, respectively, and $x_{Aa}$ takes the
 value 1 for individuals with genotype $Aa$. The variables $x_B$ and
 $x_{Bb}$ are defined similarly.  To test for epistasis, the fit of
 this model is compared to an additive model where
 $I_{AA}\!=\!I_{AD}\!=\!I_{DA}\!=\!I_{DD}\!=\!0$.  The test statistic
 for a likelihood ratio test is minus twice the difference between the
 log-likelihood of the additive and the full model.  This is
 equivalent to testing if the circuits $c_7=c_8=c_9=c_{10}=0$.

Case-control association data can be modeled using logistic regression:
\begin{eqnarray*}
 \log \left( \frac{f_{ij}}{1-f_{ij}} \right) &=&  \tilde{f} + f_a \cdot x_{A} + f_b \cdot x_{B} + \delta_a \cdot x_{Aa} + \delta_b \cdot x_{Bb}  + I_{AA} \cdot x_{A}x_{B}\\
   & & + I_{AD} \cdot x_{Aa}x_{B}  + I_{DA} \cdot x_{A}x_{Bb} + I_{DD} \cdot x_{Aa}x_{BB} + \epsilon,
 \end{eqnarray*}
 where the $f_{ij}$ are penetrances and the dummy variables $x_*$ are
 defined as above.  By using logistic regression the log-odds scale is
 chosen as the scale of interest, and additivity on that scale
 corresponds to no interaction.  A likelihood ratio test for epistasis
 compares the fit of the full model to an additive model where 
$I_{AA}\!=\!I_{AD}\!=\!I_{DA}\!=\!I_{DD}\!=\!0$.  This test is equivalent to testing
 $c_7=c_8=c_9=c_{10}=0$ where the circuits is obtained by replacing
 $f_{ij}$ with $log(f_{ij}/(1-f_{ij}))$.

 Recently \citet{Zhao2006} proposed a new test to detect unlinked
 interacting disease loci.  They use an LD based interaction measure,
 $I = h_{00}h_{11} - h_{01}h_{10}$, where $h_{ij}$ is defined as the
 penetrance of a haplotype $h_{ij}$ ($h_{00}$ is the haplotype ab,
 $h_{01}$ is $aB$, etc.).  The haplotype penetrance depends on the two
 locus penetrances as well as the allele frequencies. It is easy to
 show that the interaction measure, $I$, vanishes if
 $c_7=c_8=c_9=c_{10}=0$ when the circuits are calculated using the log
 penetrance values.  In other words, this interaction measure tests
 for multiplicative penetrances.

\begin{figure}[ht]
\begin{center}
\includegraphics[scale=0.5]{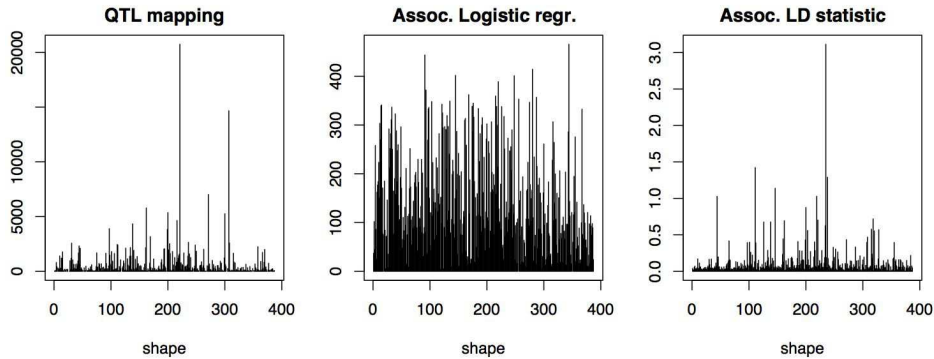}
\caption{The plots show the maximum value of the likelihood ratio test statistic observed for randomly generated data from each of the 387 shapes.}
\label{fig:power}
\end{center}
\end{figure}

We generated $45,000$ random vectors of length 9.  For the QTL
analysis we fixed the population mean of the phenotype, fixed the
allele frequencies of $A$ and $B$, and then normalized each random
vector to give the desired population mean.  For each vector we
generated 10 datasets, each with sample size 300, and fit both the
full model and an additive model.  We used the average likelihood
ratio statistic as an indicator of the power to detect interaction
for that particular model.  For each random model we then recorded
which of the 387 model shapes it fell into and for each shape looked
at maximum of the likelihood ratio statistic. In the first panel of
Figure~\ref{fig:power} we show the maximum for each shape. These maxima are highly variable between shapes, indicating that some types of
interactions are easier to detect than others.  We also observed that
there is a strong association between large values of the likelihood
ratio test statistic and the number of polygons a shape divides
the square into.

We also generated case-control data from an association study.  The
random vectors were normalized so that they all give the same
population prevalence of disease.  In the middle panel of
Figure~\ref{fig:power} we have plotted the maximum of the likelihood
ratio test statistic as a function of the shape induced by the
penetrances.  The test measures deviation from additivity on a
log-odds scale, so the difference between the different shapes is relatively
small.  When the shape is calculated based on the log-odds the results
are the same as before.  Finally, in Panel 3 of Figure~\ref{fig:power}
we plot the maximum absolute value of the interaction measure $I$.  This test
measures deviation from additivity on a log scale, yet the results
seem to be more similar to the QTL mapping case.

\section{Discussion}
The multitude of terms used to describe gene interactions are a
testament not only to the importance of describing and classifying
gene interaction, but also to the fact that even in a two-locus model
the types of interactions that can and do occur are diverse and
difficult to classify.
Most examples of gene interactions that are observed in real data do
not fall into any one of the categories typically used to describe
interactions.
Our approach overcomes this limitation
and provides a complete classification of
all two-locus models with continuous genotypic values into 69 (or 387) classes.  The shape of a
two-locus model reveals information about the types of gene
interaction present and provides a visual representation of epistasis.
By comparing an observed shape to the shapes of standard epistatic
models we see which type of interaction is strongest in the data.
Moreover, the values of the individual circuits listed in Appendix~\ref{app:circ}
provide a complete description of the epistasis in a two-locus system.
The observed shape can differ from the true underlying model shape due
to noise in the data.  Rather than assign an observation to a shape
based on the observed genotype values, one could test which circuits
are significantly different from zero and use only those circuits to
obtain the shape.

Two-locus models are frequently used to generate simulated datasets
that form the basis for studies of the power of single-locus and
two-locus methods. These can then be used e.g.\ to choose between
exhaustive two-locus searches or two-stage two-locus analyses.  There
are many examples, both for linkage analysis and association analysis,
where the results and ensuing recommendations depend on the models,
and types of gene interactions, that are considered
\citep{Marchini2005,Evans2006}.  With our complete classification it
is possible to generate data from each model class (while varying
parameters such as population prevalence and allele frequencies) and
subsequently a more thorough analysis than previously possible can be
performed.

As observed in~\citet{Carlborg2004} ``there are no striking
similarities with a Mendelian pattern of digenic epistasis'' in the
example in Section~\ref{sec:classify} and we found many types of
nontrivial interaction, including models which cannot easily be
described using existing models.  The fact that our classification is
purely mathematical lends it strength, since we can describe all
possible models and categorize them according to the relative
genotypic values.  It can easily be extended to three or more loci.
It remains to be seen whether all of the 69 types occur in nature.
Our results also provide a formalism for identifying types of
epistasis that may play a role in determining genetic variability in
populations~\citep{Turelli2006}, but we do not address these
implications in this paper.

\section{Acknowledgements}
I.B.H. was supported by grant 512066 (LSHG-CT-2004) from the European Union FP6 programme.

\bibliography{epistasis.bib}

\begin{thebibliography}{}

\bibitem[\protect\citeauthoryear{Beerenwinkel \bgroup \em et al.\egroup
  }{2006}]{Beerenwinkel2006}
N.~Beerenwinkel, L.~Pachter, and B.~Sturmfels.
\newblock Epistasis and shapes of fitness landscapes.
\newblock {\em Statistica Sinica}, in press, 2006.

\bibitem[\protect\citeauthoryear{Brem and Kruglyak}{2005}]{Brem2005}
R.~Brem and L.~Kruglyak.
\newblock The landscape of genetic complexity acreoss 5,700 gene expression
  traits in yeast.
\newblock {\em PNAS}, 102(5):1572--1577, 2005.

\bibitem[\protect\citeauthoryear{Carlborg and Haley}{2004}]{Carlborg2004b}
\"O. Carlborg and C.S. Haley.
\newblock Epistasis: too often neglected in complex trait studies?
\newblock {\em Nat. Rev. Genet.}, 5, 2004.

\bibitem[\protect\citeauthoryear{Carlborg \bgroup \em et al.\egroup
  }{2004}]{Carlborg2004}
\"O. Carlborg, P.~Hocking, D.~Burt, and C.~Haley.
\newblock Simultaneous mapping of epistatic {QTL} in chickens reveals clusters
  of {QTL} pairs with similar genetic effects on growth.
\newblock {\em Genet. Res.}, 83:197--209, 2004.

\bibitem[\protect\citeauthoryear{Cheverud and Routman}{1995}]{Routman1995}
J.M. Cheverud and J.~Routman.
\newblock Epistasis and its contribution to genetic variance components.
\newblock {\em Genetics}, 139:1455--61, 1995.

\bibitem[\protect\citeauthoryear{Clark and Wang}{1997}]{Clark1997}
A.G. Clark and L.~Wang.
\newblock Epistasis in measured genotypes: Drosophila p-element insertions.
\newblock {\em Genetics}, 147:157--63, 1997.

\bibitem[\protect\citeauthoryear{Cordell}{2002}]{Cordell2002}
H.J. Cordell.
\newblock Epistasis: what it means, what it doesn't mean, and statistical
  methods to detect it in humans.
\newblock {\em Hum. Mol. Gen.}, 11(20):2463--2468, 2002.

\bibitem[\protect\citeauthoryear{DeLoera \bgroup \em et al.\egroup }{to
  appear}]{DeLoera}
J.~A. DeLoera, J.~Rambau, and F.~Santos.
\newblock {\em Triangulations: Applications, Structures, Algorithms}.
\newblock Springer, to appear.

\bibitem[\protect\citeauthoryear{Evans \bgroup \em et al.\egroup
  }{2006}]{Evans2006}
D.M. Evans, J.~Marchini, A.P. Morris, and L.R. Cardon.
\newblock Two-stage two-locus models in genome-wide association.
\newblock {\em PLOS Genetics}, 2(9):1424--1432, 2006.

\bibitem[\protect\citeauthoryear{Fisher}{1918}]{Fisher1918}
R.A. Fisher.
\newblock The correlations between relatives on the supposition of mendelian
  inheritance.
\newblock {\em Trans. R. Soc. Edinburgh}, 52:399--433, 1918.

\bibitem[\protect\citeauthoryear{Gauderman}{2002}]{Gauderman2002}
W.J. Gauderman.
\newblock Sample size requirements for association studies of gene-gene
  interaction.
\newblock {\em Int. J. of Epidemiology}, 155(5):478--484, 2002.

\bibitem[\protect\citeauthoryear{Hirschhorn \bgroup \em et al.\egroup
  }{2002}]{Hirschhorn2002}
J.N. Hirschhorn, K.~Lohmueller, E.~Byrne, and K.~Hirschhorn.
\newblock A comprehensive review of genetic association studies.
\newblock {\em Genet. Med.}, 4:45--61, 2002.

\bibitem[\protect\citeauthoryear{Hoh and Ott}{2003}]{Hoh2003}
J.~Hoh and J.~Ott.
\newblock Mathematical multi-locus approaches to localizing complex human trait
  genes.
\newblock {\em Nature Rev. Genet.}, 4:701--709, 2003.

\bibitem[\protect\citeauthoryear{Li and Reich}{2000}]{LiReich2000}
W.~Li and J.~Reich.
\newblock A complete enumeration and classification of two-locus disease
  models.
\newblock {\em Hum Hered}, 50:334--349, 2000.

\bibitem[\protect\citeauthoryear{Marchini \bgroup \em et al.\egroup
  }{2005}]{Marchini2005}
J.~Marchini, P.~Donnelly, and L.R. Cardon.
\newblock Genome-wide strategies for detecting multiple loci that influence
  complex diseases.
\newblock {\em Nature Genetics}, 37(4):413--417, 2005.

\bibitem[\protect\citeauthoryear{Moore}{2005}]{Moore2005}
J.H. Moore.
\newblock A global view of epistasis.
\newblock {\em Nature Genetics}, 37(1), 2005.

\bibitem[\protect\citeauthoryear{P\'erez-Enciso}{2006}]{Perez2006}
M.~P\'erez-Enciso.
\newblock Multiple association analysis via simulated annealing (massa).
\newblock {\em Bioinformatics}, 22(5), 2006.

\bibitem[\protect\citeauthoryear{Phillips}{1998}]{Phillips1998}
P.C. Phillips.
\newblock The language of gene interaction.
\newblock {\em Genetics}, 149:1167--1171, 1998.

\bibitem[\protect\citeauthoryear{Rambau}{2002}]{Rambau2002}
J.~Rambau.
\newblock Topcom: Triangulations of point configurations and oriented matroids.
\newblock In {\em Proc. Int. Congress of Mathematical Software, ICMS}, 2002.

\bibitem[\protect\citeauthoryear{Storey \bgroup \em et al.\egroup
  }{2005}]{Storey2005}
J.D. Storey, J.M. Akey, and L.~Kruglyak.
\newblock Multiple locus linkage analysis of genomewide expression in yeast.
\newblock {\em PLOS Biology}, 3(8), 2005.

\bibitem[\protect\citeauthoryear{Turelli and Barton}{2006}]{Turelli2006}
M.~Turelli and H.~Barton.
\newblock Will multilocus epistasis and population bottlenecks increase
  additive genetic variance?
\newblock {\em Evolution}, 60(9):in press, 2006.

\bibitem[\protect\citeauthoryear{Wade}{2001}]{Wade2001}
M.J. Wade.
\newblock Epistasis, complex traits, and mapping genes.
\newblock {\em Genetica}, 112-113:59--69, 2001.

\bibitem[\protect\citeauthoryear{Wade}{2002}]{Wade2002}
M.J. Wade.
\newblock A gene's eye view of epistasis, selection and speciation.
\newblock {\em J. Evol. Biol.}, 15:337--346, 2002.

\bibitem[\protect\citeauthoryear{Zhao \bgroup \em et al.\egroup
  }{2006}]{Zhao2006}
J.~Zhao, L.~Jin, and M.~Xiong.
\newblock Test for interaction between two unlinked loci.
\newblock {\em Am. J. of Hum. Genet}, 79:831--845, 2006.

\bibitem[\protect\citeauthoryear{Ziegler}{1995}]{GKZ}
G.~Ziegler.
\newblock {\em Lectures on Polytopes}, volume 152.
\newblock Springer, New York, NY, 1995.

\end{thebibliography}

\newpage
\appendix
\section{Appendix A: Polyhedral subdivisions}
\label{app:tri}
Our classification is based on the theory of \begin{it}regular polyhedral subdivisions\end{it}.

\begin{definition} A \begin{it}polyhedral subdivision\end{it} of a
  point set $A$ is a decomposition of $conv(A)$, the convex hull of
  $A$, into a finite number of bounded polyhedra, such that the union
  of these polyhedra is $conv(A)$, and the intersection of any two
  polyhedra is a common face of each (possibly the empty face).
\end{definition}

A polyhedral subdivision where all the polyhedra are simplices is
called a \begin{it}triangulation\end{it}.  We can construct
a \begin{it}regular\end{it} polyhedral subdivision of a point set $A$
using the following construction: Assign to every point $a_i$ in $A$ a
`height', $h_i$.  Then lift each point in $A$ to its specified height
by forming the new point set
 $$\tilde{A}=\{(a_i,h_i)\}_{a\in A}.$$
Take $conv(\tilde{A})$, and consider its ``upper faces", that is, the faces whose outward-pointing normal vector has its last coordinate positive. Project each upper face onto $conv(A)$, by dropping the final coordinate of each point. In this manner, we obtain a polyhedral subdivision of $A$.  Note that some points of $A$ may not be used in this subdivision.

\begin{remark} In the construction of an induced subdivision there is some ambiguity as to the whether to project with the lower or upper faces of $conv(\tilde{A})$.  Both conventions are commonplace. We chose to use the upper faces in order to stay consistent with literature on induced subdivisions and gene epistasis~\citep{Beerenwinkel2006}.
\end{remark}

If the set of heights $\{h_i\}$ is sufficiently generic, then the subdivision induced by the heights will be a triangulation.  We will only consider regular subdivisions and triangulations, thus we will use the term ``subdivision" to mean ``regular polyhedral subdivision", and ``triangulation" to mean ``regular triangulation".  For more on polyhedral geometry see the book \citet{DeLoera}.


\newpage
\section{Appendix B: Circuits}
\label{app:circ}

\begin{eqnarray*}
c_{1} & = &   -2 \delta_a  + 2 I_{AD}  \\ 
c_{2} & = &   -2 \delta_a    -2 I_{DD}  \\ 
c_{3} & = &   -2 \delta_a    -2 I_{AD}  \\ 
c_{4} & = &   -2 \delta_b  + 2 I_{DA}  \\ 
c_{5} & = &   -2 \delta_b    -2 I_{DD}  \\ 
c_{6} & = &   -2 \delta_b    -2 I_{DA}  \\ 
c_{7} & = & \phantom{-} I_{AA}  + I_{AD}  + I_{DA}  + I_{DD}  \\ 
c_{8} & = & \phantom{-} I_{AA}  - I_{AD}  + I_{DA}  - I_{DD}  \\ 
c_{9} & = & \phantom{-} I_{AA}  - I_{AD}  - I_{DA}  + I_{DD}  \\ 
c_{10} & = & \phantom{-} I_{AA}  + I_{AD}  - I_{DA}  - I_{DD}  \\ 
c_{11} & = &   -2 \delta_a    -2 \delta_b  + 2 I_{AA}    -2 I_{DD}  \\ 
c_{12} & = &   -2 \delta_a    -2 \delta_b    -2 I_{AA}    -2 I_{DD}  \\ 
c_{13} & = & \phantom{-} 2 I_{AA}  + 2 I_{DA}  \\ 
c_{14} & = & \phantom{-} 2 I_{AA}    -2 I_{DA}  \\ 
c_{15} & = & \phantom{-} 2 I_{AA}  + 2 I_{AD}  \\ 
c_{16} & = & \phantom{-} 2 I_{AA}    -2 I_{AD}  \\ 
c_{17} & = & \phantom{-} 4 I_{AA}  \\ 
c_{18} & = &   -2 \delta_a  + I_{AA}  + I_{DA}  - I_{DD}  + I_{AD}  \\ 
c_{19} & = & \phantom{-} 2 \delta_a  + I_{AA}  + I_{DA}  + I_{DD}  - I_{AD}  \\ 
c_{20} & = &   -2 \delta_a  + I_{AA}  - I_{DA}  - I_{DD}  - I_{AD}  \\ 
c_{21} & = & \phantom{-} 2 \delta_a  + I_{AA}  - I_{DA}  + I_{DD}  + I_{AD}  \\ 
c_{22} & = &   -2 \delta_b  + I_{AA}  + I_{DA}  - I_{DD}  + I_{AD}  \\ 
c_{23} & = &   -2 \delta_b  - I_{AA}  + I_{DA}  - I_{DD}  - I_{AD}  \\ 
c_{24} & = &   -2 \delta_b  + I_{AA}  - I_{DA}  - I_{DD}  - I_{AD}  \\ 
c_{25} & = &   -2 \delta_b  - I_{AA}  - I_{DA}  - I_{DD}  + I_{AD}  \\ 
c_{26} & = &   -2 \delta_a  + 2 I_{AA}  \\ 
c_{27} & = & \phantom{-} 2 \delta_a  + 2 I_{AA}  \\ 
c_{28} & = &   -2 \delta_b  + 2 I_{AA}  \\ 
c_{29} & = &   -2 \delta_b    -2 I_{AA}  \\ 
c_{30} & = & \phantom{-} 2 \delta_a    -2 \delta_b  \\ 
\end{eqnarray*}

\begin{eqnarray*}
c_{31} & = & \phantom{-} 2 \delta_a  + 2 I_{AA}  + 2 I_{DA}  + 2 I_{DD}  \\ 
c_{32} & = &   -2 \delta_a  + 2 I_{AA}  + 2 I_{DA}    -2 I_{DD}  \\ 
c_{33} & = & \phantom{-} 2 \delta_a  + 2 I_{AA}    -2 I_{DA}  + 2 I_{AD}  \\ 
c_{34} & = &   -2 \delta_a  + 2 I_{AA}    -2 I_{DA}    -2 I_{AD}  \\ 
c_{35} & = &   -2 \delta_b    -2 I_{AA}    -2 I_{DD}  + 2 I_{AD}  \\ 
c_{36} & = & \phantom{-} 2 \delta_b  + 2 I_{AA}  + 2 I_{DD}  + 2 I_{AD}  \\ 
c_{37} & = &   -2 \delta_a  + 2 I_{AA}  + 2 I_{DA}  + 2 I_{AD}  \\ 
c_{38} & = & \phantom{-} 2 \delta_a  + 2 I_{AA}  + 2 I_{DA}    -2 I_{AD}  \\ 
c_{39} & = &   -2 \delta_b  + 2 I_{AA}    -2 I_{DD}  + 2 I_{AD}  \\ 
c_{40} & = & \phantom{-} 2 \delta_b  + 2 I_{AA}  + 2 I_{DA}    -2 I_{AD}  \\ 
c_{41} & = &   -2 \delta_b  + 2 I_{AA}    -2 I_{DD}    -2 I_{AD}  \\ 
c_{42} & = &   -2 \delta_b    -2 I_{AA}  + 2 I_{DA}    -2 I_{AD}  \\ 
c_{43} & = & \phantom{-} 2 \delta_a  + 2 I_{AA}    -2 I_{DA}  + 2 I_{DD}  \\ 
c_{44} & = &   -2 \delta_b  + 2 I_{AA}  + 2 I_{DA}  + 2 I_{AD}  \\ 
c_{45} & = &   -2 \delta_b  + 2 I_{AA}    -2 I_{DA}    -2 I_{AD}  \\ 
c_{46} & = &   -2 \delta_a  + 2 I_{AA}    -2 I_{DA}    -2 I_{DD}  \\ 
c_{47} & = &   -2 \delta_a  + 4 I_{AA}  + 2 I_{AD}  \\ 
c_{48} & = &   -2 \delta_b  + 4 I_{AA}    -2 I_{DA}  \\ 
c_{49} & = &   -2 \delta_a  + 4 I_{AA}    -2 I_{AD}  \\ 
c_{50} & = & \phantom{-} 2 \delta_a  + 4 I_{AA}    -2 I_{AD}  \\ 
c_{51} & = &   -2 \delta_b    -4 I_{AA}  + 2 I_{DA}  \\ 
c_{52} & = & \phantom{-} 2 \delta_a  + 4 I_{AA}  + 2 I_{AD}  \\ 
c_{53} & = & \phantom{-} 2 \delta_b  + 4 I_{AA}  + 2 I_{DA}  \\ 
c_{54} & = &   -2 \delta_b  + 4 I_{AA}  + 2 I_{DA}  \\ 
c_{55} & = &   -2 \delta_a  + 2 \delta_b  + 2 I_{DA}  + 2 I_{AD}  \\ 
c_{56} & = & \phantom{-} 2 \delta_a    -2 \delta_b  + 2 I_{DA}  + 2 I_{AD}  \\ 
c_{57} & = & \phantom{-} 2 \delta_a    -2 \delta_b  + 2 I_{DA}    -2 I_{AD}  \\ 
c_{58} & = & \phantom{-} 2 \delta_a    -2 \delta_b    -2 I_{DA}  + 2 I_{AD}  \\ 
c_{59} & = & \phantom{-} 2 \delta_a  + 2 \delta_b  + 4 I_{AA}  + 2 I_{DA}    -2 I_{AD}  \\ 
c_{60} & = &   -2 \delta_a    -2 \delta_b  + 4 I_{AA}    -2 I_{DA}    -2 I_{AD}  \\ 
\end{eqnarray*}

\begin{eqnarray*}
c_{61} & = &   -2 \delta_a    -2 \delta_b  + 4 I_{AA}  + 2 I_{DA}  + 2 I_{AD}  \\ 
c_{62} & = &   -2 \delta_a    -2 \delta_b    -4 I_{AA}  + 2 I_{DA}    -2 I_{AD}  \\ 
c_{63} & = & \phantom{-} 2 \delta_a    -4 \delta_b    -2 I_{AA}    -2 I_{DA}  + 2 I_{AD}  \\ 
c_{64} & = & \phantom{-} 4 \delta_a    -2 \delta_b  + 2 I_{AA}  + 2 I_{DA}    -2 I_{AD}  \\ 
c_{65} & = &   -4 \delta_a  + 2 \delta_b  + 2 I_{AA}  + 2 I_{DA}  + 2 I_{AD}  \\ 
c_{66} & = & \phantom{-} 2 \delta_a    -4 \delta_b    -2 I_{AA}  + 2 I_{DA}    -2 I_{AD}  \\ 
c_{67} & = & \phantom{-} 2 \delta_a    -4 \delta_b  + 2 I_{AA}    -2 I_{DA}    -2 I_{AD}  \\ 
c_{68} & = & \phantom{-} 4 \delta_a    -2 \delta_b  + 2 I_{AA}    -2 I_{DA}  + 2 I_{AD}  \\ 
c_{69} & = & \phantom{-} 2 \delta_a    -4 \delta_b  + 2 I_{AA}  + 2 I_{DA}  + 2 I_{AD}  \\ 
c_{70} & = & \phantom{-} 4 \delta_a    -2 \delta_b    -2 I_{AA}  + 2 I_{DA}  + 2 I_{AD}  \\ 
c_{71} & = & \phantom{-} 4 \delta_a    -2 \delta_b  + 4 I_{AA}  + 2 I_{DA}    -4 I_{AD}  \\ 
c_{72} & = & \phantom{-} 4 \delta_a    -2 \delta_b    -4 I_{AA}  + 2 I_{DA}  + 4 I_{AD}  \\ 
c_{73} & = & \phantom{-} 4 \delta_a    -2 \delta_b  + 4 I_{AA}    -2 I_{DA}  + 4 I_{AD}  \\ 
c_{74} & = & \phantom{-} 2 \delta_a    -4 \delta_b    -4 I_{AA}  + 4 I_{DA}    -2 I_{AD}  \\ 
c_{75} & = &   -2 \delta_a  + 4 \delta_b  + 4 I_{AA}  + 4 I_{DA}    -2 I_{AD}  \\ 
c_{76} & = &   -4 \delta_a  + 2 \delta_b  + 4 I_{AA}  + 2 I_{DA}  + 4 I_{AD}  \\ 
c_{77} & = & \phantom{-} 2 \delta_a    -4 \delta_b  + 4 I_{AA}  + 4 I_{DA}  + 2 I_{AD}  \\ 
c_{78} & = & \phantom{-} 2 \delta_a    -4 \delta_b  + 4 I_{AA}    -4 I_{DA}    -2 I_{AD}  \\ 
c_{79} & = &   -4 \delta_a    -2 \delta_b    -2 I_{DA}    -4 I_{DD}  \\ 
c_{80} & = &   -2 \delta_a    -4 \delta_b    -4 I_{DD}  + 2 I_{AD}  \\ 
c_{81} & = &   -4 \delta_a    -2 \delta_b  + 2 I_{DA}    -4 I_{DD}  \\ 
c_{82} & = &   -2 \delta_a    -4 \delta_b    -4 I_{DD}    -2 I_{AD}  \\ 
c_{83} & = &   -2 \delta_a    -2 \delta_b  + I_{AA}  + I_{DA}    -3 I_{DD}  + I_{AD}  \\ 
c_{84} & = &   -2 \delta_a    -2 \delta_b  - I_{AA}  + I_{DA}    -3 I_{DD}  - I_{AD}  \\ 
c_{85} & = &   -2 \delta_a    -2 \delta_b  + I_{AA}  - I_{DA}    -3 I_{DD}  - I_{AD}  \\ 
c_{86} & = &   -2 \delta_a    -2 \delta_b  - I_{AA}  - I_{DA}    -3 I_{DD}  + I_{AD}  \\ 
\end{eqnarray*}

\end{document}